**Interaction of Magnetic Fields with Spinons in a Fractionalized State**

Yu Zhang[1], Hengdi Zhao[1], Tristan R. Cao[1], Rahul Nandkishore[1,2], Pedro Schlottmann[3], Lance De Long[4], and Gang Cao[1,5*]

[1]*Department of Physics, University of Colorado at Boulder, Boulder, CO 80309, USA*

[2]*Center for Theory of Quantum Matter, University of Colorado at Boulder, Boulder, CO 80309, USA*

[3]*Department of Physics, Florida State University, Tallahassee, FL 32306, USA*

[4]*Department of Physics and Astronomy, University of Kentucky, Lexington, KY 40506, USA*

[5]*Center for Experiments on Quantum Materials, University of Colorado at Boulder, Boulder, CO 80309, USA*

**Abstract**

The 4d-electron trimer lattice $Ba_4Nb_{1-x}Ru_{3+x}O_{12}$ exhibits either a quantum spin liquid (QSL) or a heavy-fermion strange metal (HFSM) phase, depending on Nb content. In the QSL state, itinerant spinons act as effective heat carriers, enhancing thermal conductivity. Strikingly, applying a magnetic field up to 14 T causes an abrupt, up-to-5000% increase in heat capacity below 150 mK, disrupting the linear temperature dependence typical of both phases. Meanwhile, AC susceptibility and electrical resistivity remain nearly unchanged, while thermal conductivity drops by up to 40% below 4 K. These results suggest spinons, despite being charge-neutral, are highly sensitive to magnetic fields at low temperatures. We propose that the magnetic field could induce Anderson localization of spinons, creating emergent non-magnetic two-level systems responsible for the Schottky-like anomaly in heat capacity. These findings point to a previously unexplored regime of spinon dynamics, potentially governed by field-induced localization and distinct from conventional magnetic or transport signatures.

*Corresponding author: gang.cao@colorado.edu

## Introduction

Recent experimental work indicates that the trimer lattice $Ba_4Nb_{1-x}Ru_{3+x}O_{12}$ ($|x| < 0.20$) presents quantum spin liquid (QSL) physics, strong electron correlations tunable across a metal-to-insulator transition, and a heavy-fermion strange metal (HFSM) phase, as illustrated in Fig.1a [1]. The most striking feature shared by the entire $Ba_4Nb_{1-x}Ru_{3+x}O_{12}$ series is the *persistent linearity* of both the low-temperature heat capacity C and thermal conductivity κ. These curious behaviors are accompanied by an extraordinarily large Sommerfeld coefficient γ and exchange energy $\theta_{CW}$ as well as the *absence of magnetic order down to 50 mK,* independent of the ground state type (Fig.1a). Moreover, the charge-insulating QSL is a much better thermal conductor than the HFSM, and this is best explained by *a charge and spin separation* [1]. The key to the extraordinary behavior of this trimer lattice is a conjectured robust heavy Fermi surface of charge-neutral spinons that provides a unified framework for describing the novel phenomena observed throughout the entire series [1]. This discovery marks a breakthrough in the decades-long search for QSLs [e.g., 2-11].

Our most recent investigation of the trimer lattice $Ba_4Nb_{1-x}Ru_{3+x}O_{12}$ ($|x| < 0.20$; the sign of x can be either positive or negative) reveals an enigmatic, yet intriguing relationship between spinons and applied magnetic fields at low temperatures T, which has not yet been explored. In essence, applied magnetic field H (up to 14 T) unexpectedly breaks the signature linearity of the heat capacity C by inducing an abrupt rise in C near an onset temperature $T_s$ = 150 mK for both the QSL and HFSM. Consequently, C is strongly increased by as much as 5000% at 50 mK. *The field-enhanced change in C (ΔC) is remarkably independent of the orientation of H, indicating the absence of orbital degrees of freedom, consistent with the nature of spinons*. In contrast, no corresponding anomalies are discerned in both the AC magnetic susceptibility χ' of the QSL and HFSM, or the electrical resistivity ρ of the HFSM under the same conditions. Furthermore, application of H drastically reduces the thermal conductivity κ (particularly below 4 K) by as much as 40% for both the QSL and HFSM, but causes no discernible change in C, and therefore the density of states in the same T range (1.7 < T < 10 K). In short, application of H leads to an exotic quantum state featuring the abrupt rise in C at the most unlikely circumstance of T < 150 mK and $\mu_o H$ = 14 T. *We argue that this emergent quantum state may be a result of field-induced localization of spinons when T approaches absolute zero*.

The $Ba_4Nb_{1-x}Ru_{3+x}O_{12}$ series forms in a trimer lattice that consists of three face-sharing metal-oxygen octahedra (Fig.1a) [1]. A trimer lattice often behaves unconventionally because the internal degrees of freedom between the three metal ions the trimer octahedra provide an extra, decisive interaction among other fundamental interactions (e.g., Coulomb and spin-orbit-interactions) that dictates physical properties. It has become increasingly clear that heavy (high atomic number) trimer lattices promise a unique pathway for discoveries of new quantum states absent in materials with other types of lattices, such as triangular and perovskite lattices [1, 12-29]. Indeed, the 4*d*-electron trimer lattice is a perfect candidate for such novel behavior [1].

**Results**

The trimer lattice adopts a rhombohedral structure with the R-3 space group (No. 148), in which a Nb-O monomer separates trimer layers along the *c* axis (Fig.1a) [1, 12, 30]. Depending on Nb concentration, the system exhibits both HFSM and adjacent QSL phases [1]. The former corresponds to $Ba_4Nb_{0.81}Ru_{3.19}O_{12}$ or $Nb_{0.81}$ and the latter $Ba_4Nb_{1.16}Ru_{2.84}O_{12}$ or $Nb_{1.16}$ (Fig.1a).

***Abrupt rise in heat capacity and entropy below 150 mK and at 14 T***

The exotic behavior of the heat capacity C is the central observation of this work. We first focus on C in the temperature range 50 mK – 1 K at $\mu_o H$ = 0 and 14 T for the HFSM and QSL (Figs.1b-c). As already established in our previous study [1], at $\mu_o H = 0$ T, a linear C is persistent down to 50 mK with a large Sommerfeld coefficient $\gamma \leq 2650$ mJ/mole $K^2$ (Figs.1b-c). However, for $\mu_o H = 14$ T, C rises rapidly below $T_s = 150$ mK for both the HFSM and QSL (Figs.1b-c).

To quantify the field-induced increase in entropy, we present the following: Temperature-dependent entropy difference, ΔS(T), between 14 T and 0 T for the QSL is calculated as ΔS(T) = $\int_{T}^{0.46K} [C(14Tesla) - C(0)]/T' dT'$, where T is the argument of ΔS, and $T'$ is the integration variable, shown in Fig. 1d. The upper bound is fixed at 0.46 K because for T < 0.46 K, C(14 T) > C(0 T), indicating that the magnetic field enhances low-energy excitations in this range (Supplementary Fig.1). ΔS(T) thus represents the additional entropy induced by the magnetic field as the system is cooled from the upper bound 0.46 K down to each T, which is the lower bound of the integration. Notably, C (T>Ts) is essentially independent of magnetic field for both the QSL and HFSM phases (Figs.1b-1c and 4d-4e).

The field-induced entropy ΔS (T) = 0.22 J/mol K at 0.05 K. Compared to the maximum entropy removal of a two-level system $S = R \ln 2$, the fraction in percentage is 3.7%. This implies that a small but non-negligible population of localized spinons contributes to the entropy at 0.05 K and 14 T. This value provides a direct estimate of the entropy involved in the low-temperature upturn and suggests a significant field-induced increase in the density of accessible low-energy states. The population of localized spinons is expected to increase with further decreases of T below 0.05 K since ΔS (T) rapidly rises below Ts (Fig.1d).

Moreover, the field-induced increase in C, defined by ΔC/C(0) = [C(H) - C(0)]/C(0), can reach as high as 5000% at 50 mK (Figs.2a-2b), consistent with the enhancement in ΔS(T) below Ts (Fig.1d). All these results conflict with the conventional wisdom that a strong applied magnetic field (e.g., 14 T) coupled with low temperature (e.g., 50 mK) should strongly depress entropy by promoting order that reduces internal energy.

In sharp contrast with C (Figs.1b-c), both the *a*-axis resistivity $\rho_a$ for the HFSM and AC magnetic susceptibility $\chi_a$' for both the QSL and HFSM measured in the same temperature range display only a *featureless* response to 14 T (Figs.2c-2d and Supplementary Fig.2). Notably, ΔC is independent of the orientation of H for both HFSM and QSL, as evidenced in C(T) for $\mu_o H$ = 14 T (Figs.1b-1c) and C(H) at T = 100 mK (Fig. 2e). The lack of the H-orientation dependence indicates *an absence of the orbital coupling* and an important role of the Zeeman interaction through which the relevant degrees of freedom couple to H, which further highlights the nature of spinons as fractional excitations.

In contrast, both $\rho_a$ and $\rho_c$ at higher temperatures (e.g, 5 K) for the QSL are a strong function of H-orientation and exhibit a strong oscillatory behavior as a function of the angle between H and the applied current I, *revealing a strong orbital dependence* (Supplementary Fig.3). The magneto-resistivity ratio, defined by $\Delta\rho/\rho(0) = [\rho(H)-\rho(0)]/\rho(0)$, can be as high as 60% at 9 T. It is remarkable that $\Delta\rho/\rho(0)$ is predominantly positive for the *a*-axis $\rho_a$ and negative for the *c*-axis $\rho_c$ (Supplementary Fig.3). Such a giant, anisotropic oscillatory magnetoresistance suggests an orbital quantum interference in the variable range hopping regime [31 and references therein], which is interesting in its own right (note no long-range magnetic order down to 50 mK). This contrasting transport behavior further highlights the spin-charge separation in the system at low temperatures.

***Absence of conventional Schottky effect***

For comparison and contrast, the heat capacity C of a related trimer metal, 9R-$BaRuO_3$, as well as insulating $Nb_2O_5$ is also measured as functions of T (Fig.1e) and H (Fig.2f) under the similar experimental conditions. Both the T- and H-dependences of C exhibit no similarities to those of the HFSM and QSL (Figs.1b-1c and 2e), which decisively rules out any possible spurious effects from the Ba, Ru, and Nb starting materials and/or contributions from the nuclear heat capacity of those elements. Therefore, *the observed $\Delta C$ anomalies must be unique to $Ba_4Nb_{1-x}Ru_{3+x}O_{12}$*. We note that the absence of unpaired electrons in $Nb_2O_5$ ($Nb^{5+}$, $4d^0$) as well as the weak electric field gradient interaction makes a conventional Schottky anomaly unlikely and explains the featureless behavior in Fig.1e, as supported by prior work [32]. Moreover, the finite value of C(T) for $BaRuO_3$ at H = 0 and low temperatures signals an emergent phase below 0.3 K.

Nevertheless, the field-induced upturn in C (Figs.1b-1c) could be related to a possible Schottky-like effect due to a splitting of two-levels, δ. This could lead to a high-T tail of a Schottky peak located well below 50 mK, therefore experimentally inaccessible; however, applied field broadens δ and shifts up the Schottky peak to higher T. As such, the Schottky contribution to C is expectedly proportional to $DT^{-2}$ (D = Schottky coefficient) for T >> δ. Combining $DT^{-2}$ with a $\gamma T$ contribution due to spinons yields a total $C = \gamma T + DT^{-2}$ that describes C(T) for both phases over 50 mK ≤ T ≤ 1 K in the presence of a strong H. *Note that the phonon contribution to C ($\sim T^3$) is either zero or negligible as the observed C show a robust linear temperature dependence over 50 mK ≤ T ≤ 8 K at H = 0* (Fig.1b-1c and [1]). This is because the phonon contribution (positive $\beta$) is compensated by the second term of the Sommerfeld expansion of the electronic contribution (negative $\beta$) yielding a measured $\beta$ that is essentially zero [1]. Fitting C(T, 14T) to $C/T = \gamma + DT^{-3}$ generates a linear fit that determines values of D for both phases at $\mu_o H$ = 14 T (Fig.3a). Additionally, C(H) basically scales with $H^2$ below 200 mK (Fig.3b). A few features are worth noting: (1) The D values for both phases are essentially identical, *suggesting that the degrees of freedom responsible for $\Delta C$ are the same*; and (2) The magnitude of the D values, $10^{-3}$ JK/mole, is three orders of magnitude greater than $10^{-6}$ JK/mole due to the quadrupolar and/or magnetic spin splitting of Ru nuclei [33 and references therein], suggesting that the Schottky physics (if such it is) is *not of nuclear origin.*

For comparison, similar measurements are conducted for a 2.5% Pr doped isostructural $Ba_4Nb_{1-x}Ru_{3+x}O_{12}$ i.e., $Ba_{3.90}Pr_{0.10}Nb_{0.84}Ru_{3.16}O_{12}$, in which the Pr doping introduces a conventional Schottky anomaly characterized by an upturn in C at H = 0 below 200 mK (Fig.3e). This clear Schottky anomaly rapidly shifts to higher T with increasing H. In contrast, the upturn in C for both the HFSM and QSL is steeper and relatively less field-sensitive, as shown in Figs.3c-3d.

We note that similar low-temperature upturns in heat capacity have been reported in elemental niobium under magnetic fields [34]. However, elemental Nb is a metallic superconductor with very different electronic structure and nuclear environment compared to the trimer-based system with $Nb^{5+}(4d^0)$ studied here. A detailed comparison with this work, including magnitude, temperature scale, and Nb content dependence, is presented in the Discussion section. As shown there, the observed anomaly in $Ba_4Nb_{1-x}Ru_{3+x}O_{12}$ is inconsistent with a nuclear Schottky origin.

All in all, the data in Fig. 3 suggest that a conventional Schottky effect alone does not provide an adequate explanation of the ΔC(H) data for $T< T_s = 150$ mK.

***Reduction of thermal conductivity under magnetic fields***

We now turn to the data for the thermal conductivity κ over 1.7 K – 10 K at selected H. Our previous study [1] has already established that the QSL is a much better thermal conductor than the HFSM and that both are dominated by spinons at low T. In this study, we observe that application of a magnetic field readily suppresses κ in both phases (Figs.4a-4b), which is inconsistent with experimental precedents [e.g., 35, 36, 37, 38, 39]. Generally, κ is proportional to C, the velocity *v* of heat carriers and the mean free path *l* of the heat carriers, i.e., κ ~ C*vl*. Because *v* and *l* are essentially constant at low T [40], C is expected to dictate κ.

In the present study, C for either phase does not change with H in the same temperature range, 1.7 K – 10 K, as shown in Fig.4d-4e. Therefore, the observed reduction of κ could indicate a significant reduction of the spinon velocity *v* due to H although a reduction of *l* of the spinons cannot be ruled out. As shown in Fig.4a, the magnetic field effect on the *a*-axis $\kappa_a$ is strong initially when H increases from 0 T to 7 T but becomes weaker as H further increases from 7 T to 14 T, suggesting a trend for saturation with increasing H, which is consistent with the above argument. Furthermore, a close examination of κ at selected H in Fig.4a-4b reveals that the magnetic-field effect on $\kappa_a$ is a strong function of T below 4 K.

We define a magneto-thermal-conductivity ratio, $\Delta\kappa_a/\kappa_a(0T) = [\kappa_a(14T) - \kappa_a(0T)]/\kappa_a(0T)$, to quantify the reduction in $\kappa_a$ due to the applied 14 T. Note that $\Delta\kappa_a$ reflects contributions from heat carriers that are susceptible to H (i.e., H-reduced $\kappa_a$). Since the QSL is a charge insulator, $\Delta\kappa_a$ must be due primarily to spinons at low T. *It is important to point out that a sizable phonon contribution to κ ($\sim T^3$) is unlikely below 8 K where both κ and C vary linearly with T* (Figs 1b-1c, 4a-4b and [1]). As shown in Fig.4c, $\Delta\kappa_a/\kappa_a(0T)$ as a function of T shows an unusually large reduction in $\kappa_a$. This reduction is considerably stronger in the QSL than in the HFSM (Fig.4c).

Such a difference is consistent with there being more spinons as heat carriers in the QSL than in the HFSM (consistent also with the larger C and κ in the QSL [1]), which naturally makes the QSL a much better thermal conductor than the HFSM (Fig.4a-4b). Indeed, ΔC/C(0) is larger in the QSL than in the HFSM (Figs.2a-2b). These observations provide an additional, key testament to the crucial role of spinons in the behavior of the trimer systems. We note that the spinons also dominate C(T) and κ(T) in the HFSM where the Wiedemann-Franz law is strongly violated, consistent with the spin-charge separation [1].

In addition, $\Delta\kappa_a/\kappa_a$(0T) for both phases exhibits a rapid downturn at T < 4 K, and a reduction in $\kappa_a$ by as much as 40% near 1.7 K (Fig.4c). Because C in the same temperature range remains unchanged (Figs.4d-4e), the increasingly negative $\Delta\kappa_a/\kappa_a$(0T) with decreasing T (Fig.4c) forcefully indicates that *the mobility of spinons decreases rapidly with decreasing T as a result of the applied magnetic field*. Indeed, it is conceivable that strong magnetic fields comparable to 14 T could eventually localize otherwise itinerant spinons at milli-Kelvin temperatures. This point is schematically illustrated in Fig.4f.

**Discussion**

The most striking experimental feature in this work is the dramatic upturn in C(T), appearing below $T_S \approx 150$ mK under a magnetic field of 14 T (Figs.1b-1c). This anomaly is independent of field orientation and is not accompanied by corresponding features in either the magnetic susceptibility or electrical resistivity, which rules out conventional phase transitions (Figs.2c-2d and Supplementary Fig.2). The scaling of $C(H) \propto H^2/T^2$ suggests a Schottky-like behavior, but with magnitudes far exceeding nuclear contributions (Fig.3a-3b). All these phenomena underscore an extraordinary susceptibility of itinerant spinons to applied magnetic fields at milli-Kevin temperatures, which is a regime not extensively explored experimentally or theoretically. The strong linear temperature dependence of C, ρ, and κ at low temperatures effectively rules out impurity or chemical disorder as contributing factors to the observed novel phenomena.

We propose that the dramatic low-T upturn in C(T) arises from the *field-induced localization of spinons,* which are itinerant and charge-neutral excitations dominating both C and κ in the QSL and HFSM. In the absence of magnetic field, the spinons are weakly antilocalized due to symplectic symmetry [41]. Application of a magnetic field breaks time-reversal symmetry and changes the symmetry class, suppressing antilocalization and enabling Anderson localization at low temperatures, as schematically shown in Fig.4g.

While the low-temperature upturn in C(T) exhibits a Schottky-like functional form, the absence of a peak near the expected Zeeman energy scale (~10 K at 14 T) indicates that the relevant two-level systems are not conventional magnetic doublets (i.e., up/down spin states). Instead, we conjecture that the observed behavior originates from spinons that become Anderson localized under strong magnetic field. In this regime, the two-level systems could arise from pairs of spatially distinct, localized spinon wavefunctions with the same spin quantum number**,** rather than from spin degenerate states. While this interpretation remains speculative, it naturally explains the emergence of a Schottky-like anomaly at ultra-low temperatures without a corresponding high-temperature peak. Indeed, our fits to $C/T = \gamma + D/T^3$ yield $D \approx 10^{-3}$ J·K/mol**,** which suggests a level splitting δ on the order of ~ 0.1 K**,** i.e., ~10 μeV (Fig.3a).

This scenario is further supported by the thermal conductivity data: while C remains constant between 2–10 K (Figs.4d-4e), κ is strongly suppressed by field (Figs.4a-4c), implying a reduction in spinon mobility. The downturn in Δκ/κ(0) below 4 K and its increasing suppression with field align with the onset of spinon localization. Importantly, this localization appears to be a crossover phenomenon: κ begins to decrease with H above 4 K, while C remains unchanged until sharply rising below $T_S \approx 150$ mK. This behavior is consistent with progressively suppressed mobility of spinons above $T_S$ and their full localization at lower T.

To specifically address the concern raised about a nuclear origin, we emphasize the following distinctions. First, the observed anomaly in $Ba_4Nb_{1-x}Ru_{3+x}O_{12}$ is three orders of magnitude stronger than typical nuclear Schottky effects, including those observed in elemental Nb [34]. Second, elemental Nb is a metallic superconductor with very different electronic structure and field response compared to our trimer-based, correlated system. The Nb ions in the trimer system are $Nb^{5+}$ ($4d^0$), that is, there are no unpaired electrons. The absence of unpaired electrons significantly reduces the magnetic field at the nucleus, which is the primary driver for nuclear spin splitting and the Schottky effect. Most critically, in our case the anomaly is independent of Nb content across a doping range where Ru/Nb ratio varies substantially, strongly arguing against a nuclear quadrupolar origin which should scale with Nb concentration. Our direct comparison with $Nb_2O_5$ and $BaRuO_3$ (Fig. 1e) further supports that the anomaly is intrinsic to the trimer lattice.

We also note that although the upturn resembles the high-temperature tail of a Schottky anomaly, the absence of a Zeeman-scale peak and the unique magnitude and field-dependence of the D coefficient (Figs. 3a-3b) remain incompatible with known nuclear effects. In short, these observations favor a novel, electronic mechanism distinct from elemental Nb behavior.

An alternative, more exotic explanation invokes fractional excitations with restricted mobility (fractons) which pair into mobile spinons in the absence of field [8, 42]. Under magnetic field, such composites could unbind into immobile constituents, but this scenario lacks a natural explanation for the sharp onset temperature $T_S$ and thus appears less consistent with the full range of observed phenomena.

Overall, our data support the **plausible conjecture** that strong magnetic fields induce localization of spinons at ultra-low temperatures, giving rise to emergent, non-magnetic low-energy degrees of freedom. This study uncovers a previously unexplored regime in quantum matter governed by the interplay between spin-charge separation and magnetic field effects.

This work suggests a potentially novel relationship between spinons and magnetic fields: as temperature approaches absolute zero, strong magnetic fields may fundamentally alter spinon behavior, leading to emergent low-energy phenomena not captured by conventional models. While the precise microscopic origin remains to be fully understood, our findings point to a regime where spin-charge separation, localization, and field-tunable dynamics converge, which warrants further investigation through both experimental and theoretical efforts.

## Methods

### *Crystal growth and characterization*

Single crystals of $Ba_4Nb_{1-x}Ru_{3+x}O_{12}$ (|x| < 0.20) were grown via a high-temperature flux method. Stoichiometric amounts of high-purity $BaCO_3$, $Nb_2O_5$, and $RuO_2$ powders were thoroughly mixed with an excess of $BaCl_2$ flux (mass ratio ≈ 1:3). The mixture was loaded into an alumina crucible and heated to 1280°C for 38 hours. The melt was then slowly cooled to 800°C at a rate of 2°C/hour, followed by rapid quenching to room temperature.

The resulting single crystals of $Ba_4Nb_{1-x}Ru_{3+x}O_{12}$ were characterized using a Bruker Quest ECO single-crystal diffractometer with an Oxford Cryosystem providing sample temperature environments ranging from 80 K to 400 K. Chemical analyses of the samples were performed using a combination of a Hitachi MT3030 Plus scanning electron microscope and an Oxford Energy Dispersive X-Ray Spectrometer (EDX). The measurements of the electrical resistivity, Hall effect, heat capacity, thermal conductivity and AC magnetic susceptibility were carried out using a Quantum Design (QD) Dynacool PPMS system with a 14-Tesla magnet, a dilution refrigerator, a homemade probe for thermal conductivity, and a set of external meters that measure current and voltage with high precision.

The samples used in this study were from the same batches as those reported in Ref. [1], with additional crystals grown for reproducibility checks. All heat capacity and thermal conductivity results shown here were confirmed on at least two independently prepared crystals per composition.

The sample mass of $Nb_2O_5$ is 5.04 mg. In the temperature interval of 0.05 – 1 K, 160 data points were taken, each temperature point was repeated twice, thus i.e., ~ 0.012 K/point. The Cal Correction value, a calibration factor, was 1. The value of 1 indicate that no correction is needed, meaning that the PPMS is directly providing accurate heat capacity values without any scaling or adjustment and that the thermal coupling is adequate.

## DATA AVAILABILITY

The data that support the findings of this work are available from the corresponding author upon request.

## ACKNOWLEDGEMENTS

G.C. thanks Xi Dai, Tai-Kai Ng, Feng Ye, Sandeep Sharma, Minhyea Lee and Longji Cui for useful discussions. Experimental work is supported by National Science Foundation via Grant No. DMR 2204811. Theoretical work by R.N. was supported by the U.S. Department of Energy (DOE), Office of Science, Basic Energy Sciences (BES) under Award # DE-SC0021346.

## AUTHOR CONTRIBUTIONS

Y.Z. conducted measurements of the physical properties; H.D.Z characterized the crystal structure of the crystals and conducted measurements of the physical properties. T.R.C. grew the single crystals and conducted measurements of the physical properties. R.N. conducted the theoretical analysis and contributed to paper revisions; P.S. conducted the theoretical analysis and contributed to paper revisions; L.D. contributed to paper revisions. G.C. initiated and directed this work, grew the single crystals, conducted experiments of physical properties, analyzed the data, constructed the figures, and wrote the paper.

## COMPETING INTERESTS

The authors declare no competing interests. Author GC serves as Guest Editor of the Special Collection on Quantum Materials Synthesis for *npj Quantum Materials* and had no role in the peer-review or decision to publish this manuscript.

**FIGURE LENGENDS**

**Fig. 1. Key thermodynamic response of the trimer lattice $Ba_4Nb_{1-x}Ru_{3+x}O_{12}$. a,** Schematic phase diagram as a function of Nb content x, illustrating the heavy spinon Fermi surface

underpinning both the HFSM and QSL phases. The extraordinarily large Sommerfeld coefficient γ (blue) and exchange energy $\theta_{CW}$ (red, right scale) are shown [1]. Inset: The crystal structure of the trimer lattice. Note that the heavy-fermion strange metal (HFSM, 1-x = 0.81 or $Nb_{0.81}$) and the insulating quantum spin liquid (QSL, 1-x = 1.16 or $Nb_{1.16}$) are adjacent to each other. **b-c**, Heat capacity C(T) for the HFSM and QSL, respectively, at 0 T and 14 T for both field orientations and for 50 mK ≤ T ≤ 1 K. The black arrows mark the onset of the upturn in C, $T_S$. **d**, Temperature-dependent entropy difference, ΔS(T), between 14 T and 0 T for the QSL, calculated as ΔS(T) = $\int_T^{0.46K} [C(14T) - C(0)]/T'dT'$ , where T is the argument of ΔS, i.e. the lower bound of the integration that varies, and $T'$ is the integration variable. The upper bound of the integration is set at 0.46 K. **e,** Comparison with 9R-$BaRuO_3$ and $Nb_2O_5$ shows no upturn in C, confirming the intrinsic nature of the effect in $Ba_4Nb_{1-x}Ru_{3+x}O_{12}$. Note that the anomaly in $BaRuO_3$ at 0.3 K may signal an emergent state.

**Fig. 2. Contrasting responses in thermodynamic and transport properties. a-b,** Relative change in heat capacity [C(H) - C(0)]/C(0) vs. H for the QSL and HFSM, respectively, showing up to 5000% enhancement at 50 mK under 14 T. Note that the QSL shows a stronger change. **c-d,** Temperature dependence of the *a*-axis $\rho_a$ for HFSM and the *a*-axis AC susceptibility $\chi'_a$ for the QSL at representative H, respectively, for 50 mK ≤ T ≤ 1 K. **e,** The field dependence of C at T = 100 mK for the QSL at both H || *a* axis and H || *c* axis, confirming spinon Zeeman coupling without orbital contribution. **f**, No anomaly is observed in $BaRuO_3$ under similar conditions, further excluding a conventional nuclear or elemental origin.

**Fig. 3. Analysis of the low-temperature heat capacity anomaly. a,** C/T vs. $1/T^3$ for 50 mK ≤ T ≤ 1 K for both the HFSM and QSL. Note the slope D (~ $10^{-3}$ JK/mole) for both phases is essentially the same, and three orders of magnitude larger than that due to nuclear contributions. **b**, C as a function of $H^2$ at representative T for the QSL. **c–d,** Upturns in C for the HFSM and QSL grow in magnitude with field but share a common onset near $T_S$ = 150 mK. **e,** In contrast, Pr-doped

$Ba_4Nb_{1-x}Ru_{3+x}O_{12}$ exhibits a conventional Schottky anomaly with stronger field sensitivity, highlighting the distinct mechanism in the undoped system.

**Fig. 4. Thermal conductivity κ and spinon dynamics under magnetic fields. a-b,** The *a*-axis thermal conductivity *a*-axis $\kappa_a$ for the QSL and HFSM under various magnetic fields. **c,** The magneto-thermal-conductivity ratio $\Delta\kappa_a/\kappa_a(0T)$ as a function of T for the QLS and HFSM where $\Delta\kappa_a = \kappa_a(14T) - \kappa_a(0T)$. Note $\Delta\kappa_a/\kappa_a(0T)$ shows up to 40% suppression near 1.7 K. **d-e**, C(T) over the same T range at $\mu_oH = 0$ and 14 T for the QSL and HFSM, respectively. Note that C(T) remains essentially unchanged at 14 T. **f,** Schematic: Magnetic field reduces spinon velocity *v* at higher T and induces localization below $T_S$. **g,** Proposed T-H phase diagram for spinon dynamics, indicating a crossover from itinerant to localized behavior.

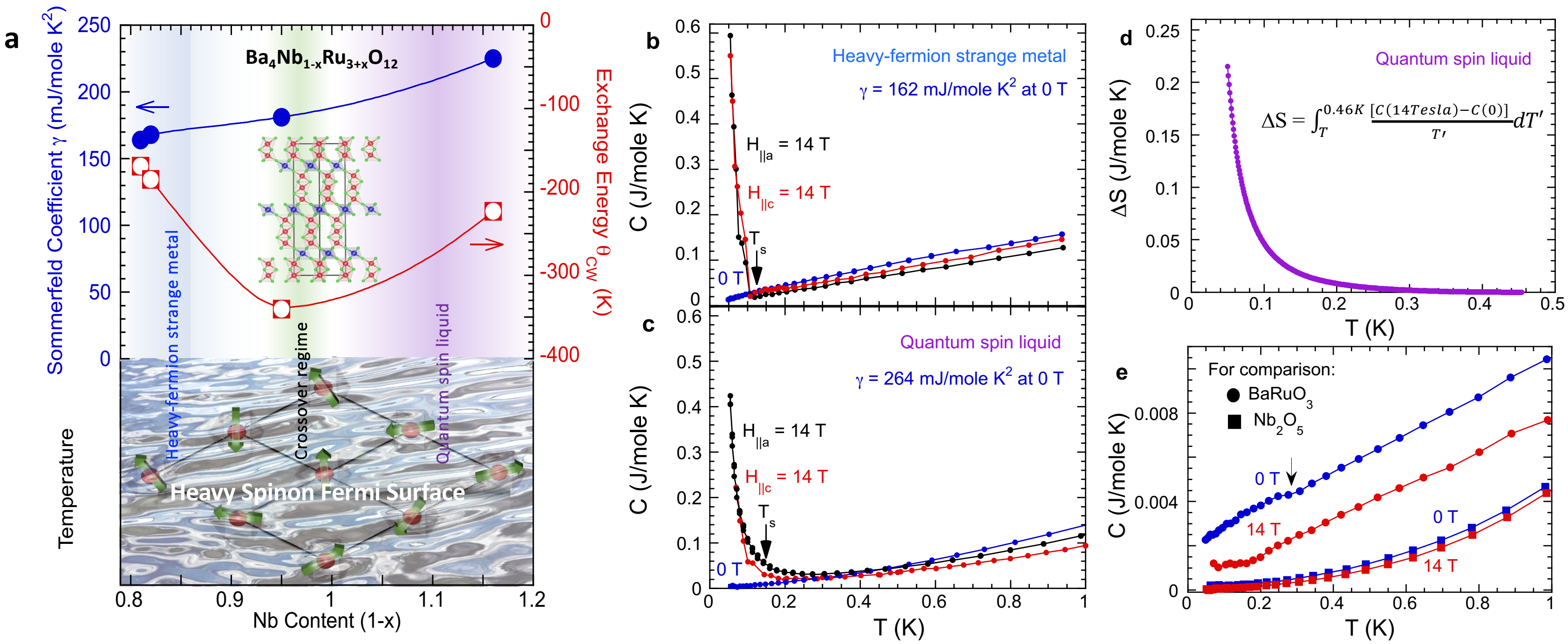
a
Ba4Nb1-xRu3+xO12
Sommerfeld Coefficient γ (mJ/mole K2)
Exchange Energy θCW (K)
Heavy-fermion strange metal
Crossover regime
Quantum spin liquid
Heavy Spinon Fermi Surface
Temperature
Nb Content (1-x)
b
Heavy-fermion strange metal
γ = 162 mJ/mole K2 at 0 T
H||a = 14 T
H||c = 14 T
Ts
0 T
C (J/mole K)
c
Quantum spin liquid
γ = 264 mJ/mole K2 at 0 T
H||a = 14 T
H||c = 14 T
Ts
0 T
C (J/mole K)
T (K)
d
Quantum spin liquid
ΔS = ∫T0.46K [C(14Tesla)−C(0)]/T′ dT′
ΔS (J/mole K)
T (K)
e
For comparison:
BaRuO3
Nb2O5
0 T
14 T
0 T
14 T
C (J/mole K)
T (K)

Figure 1

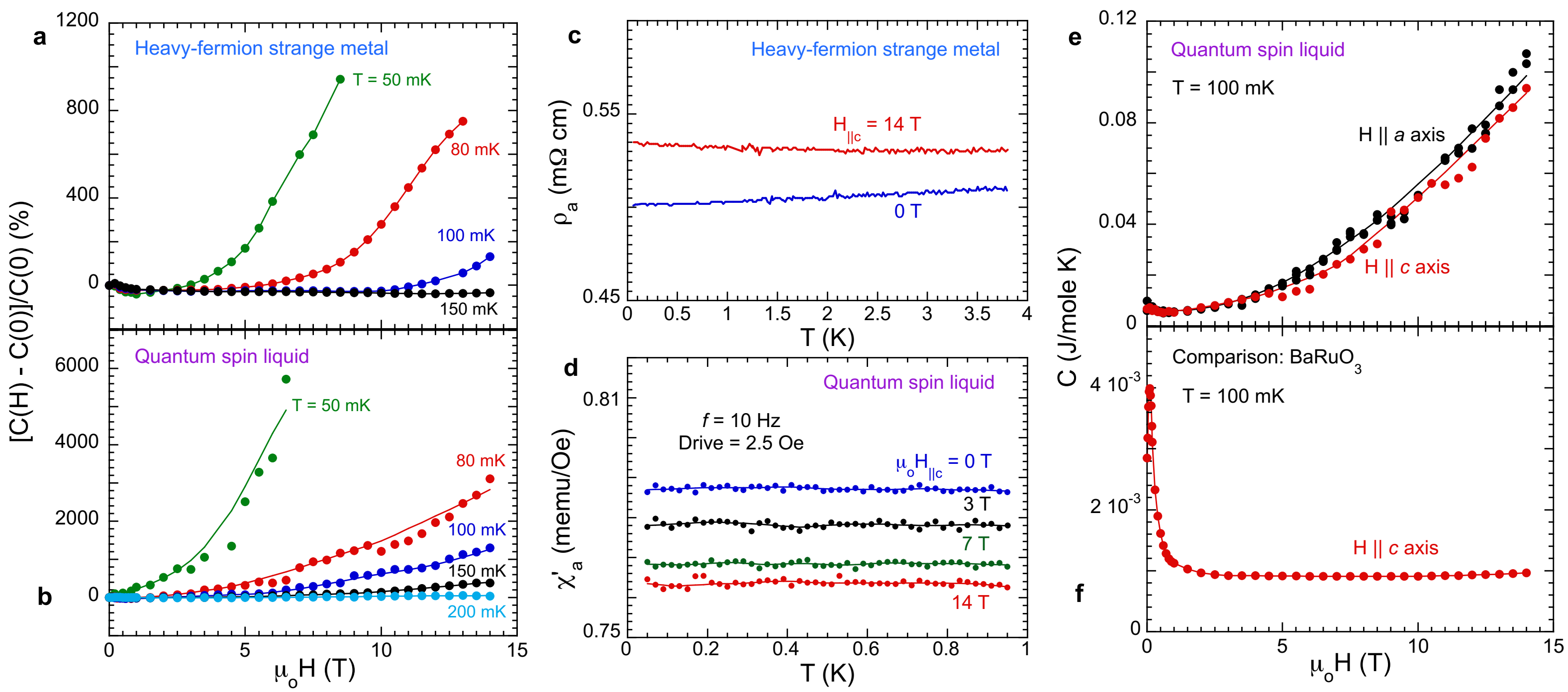
a
Heavy-fermion strange metal
T = 50 mK
80 mK
100 mK
150 mK
b
Quantum spin liquid
T = 50 mK
80 mK
100 mK
150 mK
200 mK
[C(H) - C(0)]/C(0) (%)
μoH (T)
c
Heavy-fermion strange metal
H||c = 14 T
0 T
ρa (mΩ cm)
T (K)
d
Quantum spin liquid
f = 10 Hz
Drive = 2.5 Oe
μoH||c = 0 T
3 T
7 T
14 T
χ'a (memu/Oe)
T (K)
e
Quantum spin liquid
T = 100 mK
H || a axis
H || c axis
f
Comparison: BaRuO3
T = 100 mK
H || c axis
C (J/mole K)
μoH (T)

Figure 2

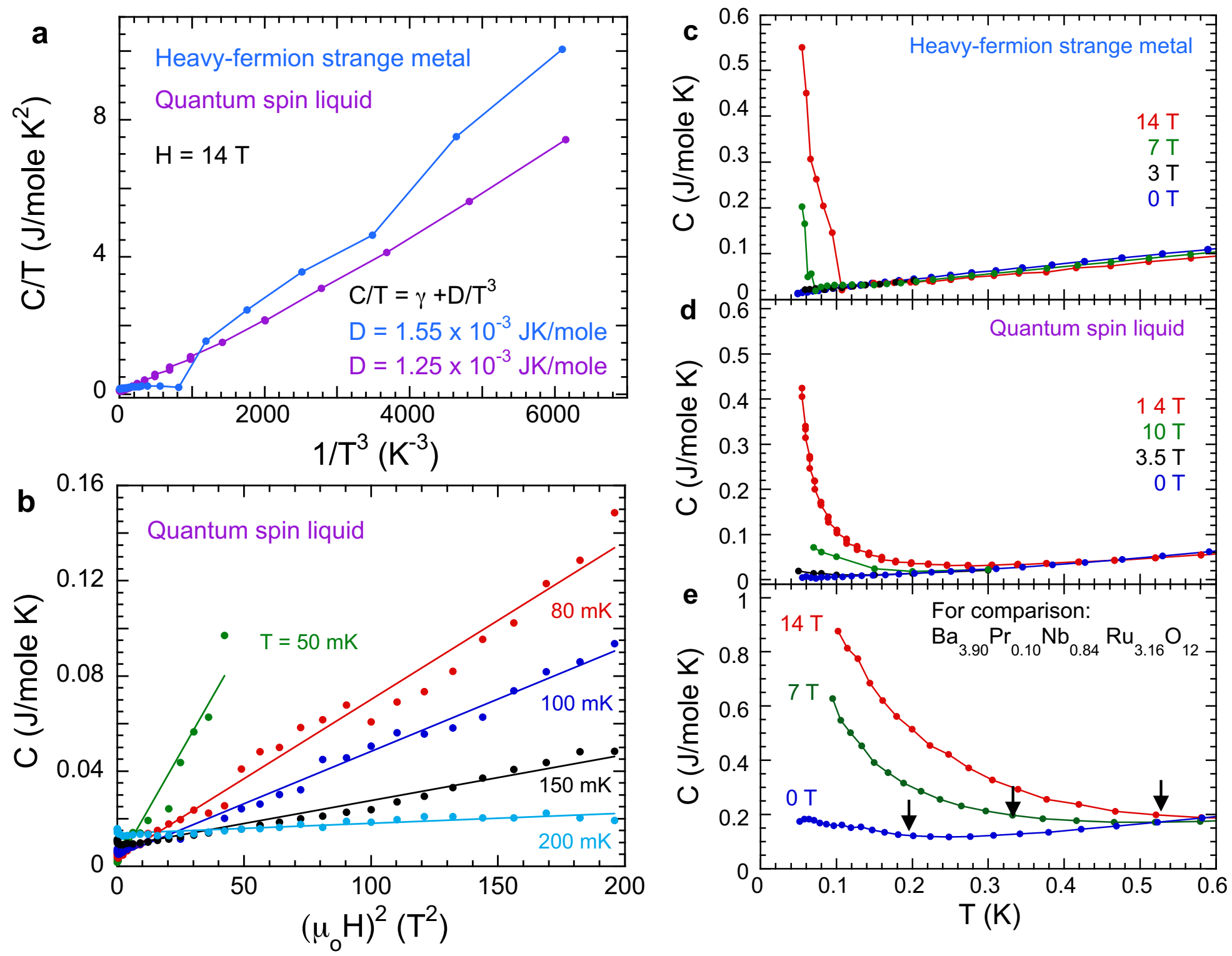
a
Heavy-fermion strange metal
Quantum spin liquid
H = 14 T
C/T = γ +D/T^3
D = 1.55 x 10^-3 JK/mole
D = 1.25 x 10^-3 JK/mole
C/T (J/mole K^2)
1/T^3 (K^-3)
b
Quantum spin liquid
T = 50 mK
80 mK
100 mK
150 mK
200 mK
C (J/mole K)
(μ_oH)^2 (T^2)
c
Heavy-fermion strange metal
14 T
7 T
3 T
0 T
d
Quantum spin liquid
1 4 T
10 T
3.5 T
0 T
e
For comparison:
Ba3.90Pr0.10Nb0.84Ru3.16O12
14 T
7 T
0 T
T (K)

Figure 3

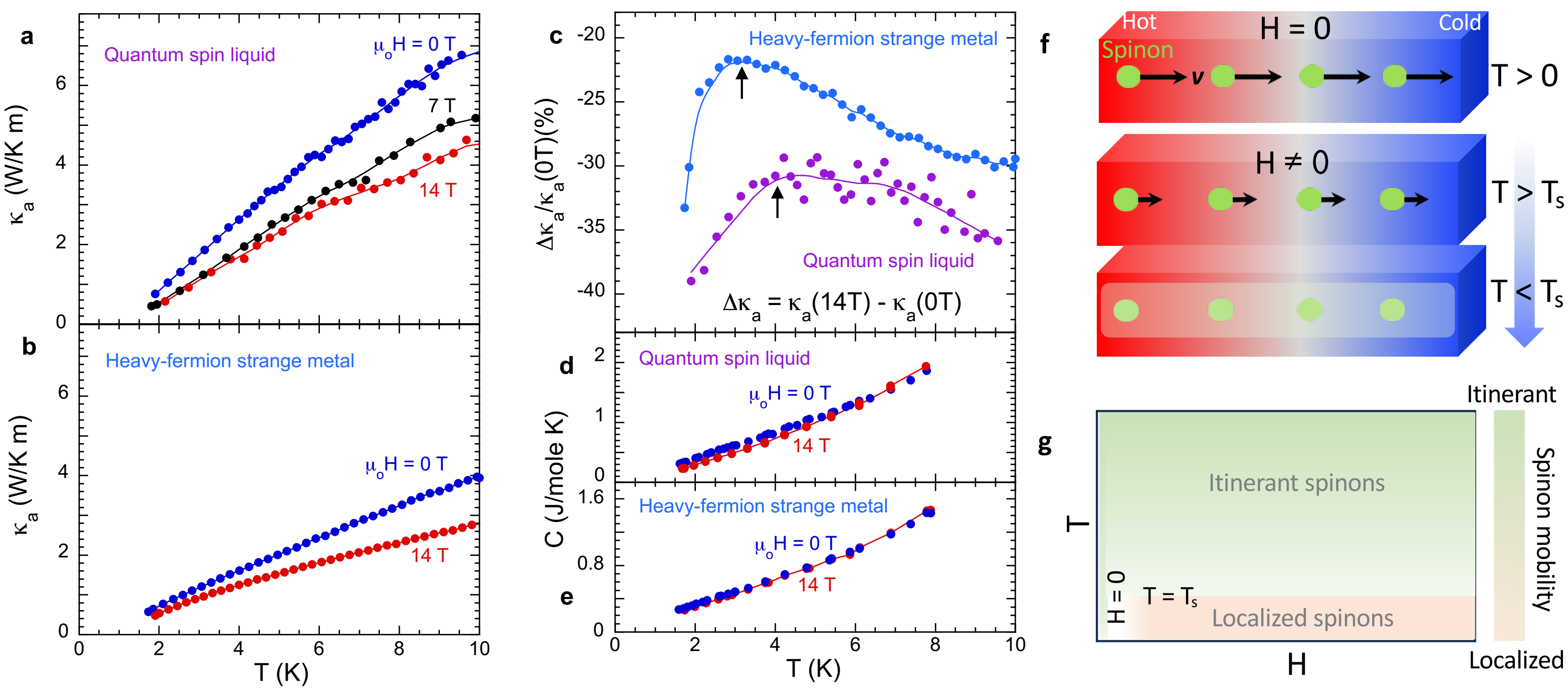

a
Quantum spin liquid
μoH = 0 T
7 T
14 T
κa (W/K m)
b
Heavy-fermion strange metal
μoH = 0 T
14 T
κa (W/K m)
T (K)
c
Heavy-fermion strange metal
Quantum spin liquid
Δκa/κa(0T)(%)
Δκa = κa(14T) - κa(0T)
d
Quantum spin liquid
μoH = 0 T
14 T
C (J/mole K)
e
Heavy-fermion strange metal
μoH = 0 T
14 T
T (K)
f
Hot
Cold
Spinon
H = 0
v
T > 0
H ≠ 0
T > Ts
T < Ts
g
Itinerant
Itinerant spinons
Spinon mobility
T
H = 0
T = Ts
Localized spinons
H
Localized

Figure 4

SUPPLEMENTAL MATERIAL

# Interaction of Magnetic Fields with Spinons in a Fractionalized State

Yu Zhang[1], Hengdi Zhao[1], Tristan R. Cao[1], Rahul Nandkishore[1,2], Pedro Schlottmann[3], Lance De Long[4], and Gang Cao[1,5*]

*[1]Department of Physics, University of Colorado at Boulder, Boulder, CO 80309, USA*

*[2]Center for Theory of Quantum Matter, University of Colorado at Boulder, Boulder, CO 80309, USA*

*[3]Department of Physics, Florida State University, Tallahassee, FL 32306, USA*

*[4]Department of Physics and Astronomy, University of Kentucky, Lexington, KY 40506, USA*

*[5]Center for Experiments on Quantum Materials, University of Colorado at Boulder, Boulder, CO 80309, USA*

## I. Experimental Details

Single crystals of $Ba_4Nb_{1-x}Ru_{3+x}O_{12}$ were grown using a flux method. Measurements of crystal structures were performed using a Bruker Quest ECO single-crystal diffractometer with an Oxford Cryosystem providing sample temperature environments ranging from 80 K to 400 K. Chemical analyses of the samples were performed using a combination of a Hitachi MT3030 Plus scanning electron microscope and an Oxford Energy Dispersive X-Ray Spectrometer (EDX). The measurements of the electrical resistivity, Hall effect, heat capacity, thermal conductivity and AC magnetic susceptibility were carried out using a Quantum Design (QD) Dynacool PPMS system with a 14-Tesla magnet, a dilution refrigerator, a homemade probe for thermal conductivity, and a set of external meters that measure current and voltage with high precision.

## II. Crystal Structure Details

The series of $Ba_4Nb_{1-x}Ru_{3+x}O_{12}$ (|x| < 0.20; the sign of x can be either positive or negative) adopts a rhombohedral structure with the R-3 space group (No. 148). Refinement details of each compound are shown in **Supplementary Tables 1-3**, respectively.

**Supplementary Table 1.** Structural parameters for single crystal $Ba_4Nb_{0.81}Ru_{3.19}O_{12}$ (**$Nb_{0.81}$**) under space group *R-3* (No. 148) at 100K. Data obtained from single crystal x-ray diffraction.

| T = 100 K | $a = b$ = 5.7322(2) Å, $c$ = 28.528(1) Å, $Z$ = 3, V = 811.79(6) Å$^3$. The agreement factor $R_1$ = 1.89% was achieved by using 1144 unique reflections with I > 4σ and a resolution of $d_{min}$ = 0.55 Å. Anisotropic atomic displacement parameters were used for all elements. | | | | | |
|---|---|---|---|---|---|---|
| | x | y | z | Occupancy | Ueq(Å$^2$) | site |
| Ba1 | 0.66667 | 0.33333 | 0.70538(2) | 1 | 0.00352(6) | 6c |
| Ba2 | 0.00000 | 0.00000 | 0.78716(2) | 1 | 0.00280(5) | 6c |
| Nb1 | 0.66667 | 0.33333 | 0.83333 | 0.71(2) | 0.0012(1) | 3b |
| Ru1 | 0.66667 | 0.33333 | 0.83333 | 0.29(2) | 0.0012(1) | 3b |
| Ru2 | 0.33333 | 0.66667 | 0.66667 | 0.94(2) | 0.00180(9) | 3a |
| Nb2 | 0.33333 | 0.66667 | 0.66667 | 0.06(2) | 0.00180(9) | 3a |
| Ru3 | 0.33333 | 0.66667 | 0.75513(2) | 0.98(2) | 0.00190(9) | 6c |
| Nb3 | 0.33333 | 0.66667 | 0.75513(2) | 0.02(2) | 0.00190(9) | 6c |
| O6 | 0.1766(3) | 0.3531(3) | 0.71086(6) | 1 | 0.0041(2) | 18f |
| O7 | 0.5027(3) | 0.4972(3) | 0.79389(6) | 1 | 0.0042(2) | 18f |

**Supplementary Table 2.** Structural parameters for single crystal $Ba_4Nb_{0.95}Ru_{3.05}O_{12}$ (**$Nb_{0.95}$**) under space group *R-3* (No. 148) at 100 K. Data obtained from single crystal x-ray diffraction.

| T = 100 K | $a = b$ = 5.7371(1) Å, $c$ = 28.533(1) Å, $Z$ = 3, V = 813.32(4) Å$^3$. The agreement factor $R_1$ = 1.31% was achieved by using 1135 unique reflections with I > 4σ and a resolution of $d_{min}$ = 0.55 Å. Anisotropic atomic displacement parameters were used for all elements. | | | | | |
|---|---|---|---|---|---|---|
| | x | y | z | Occupancy | Ueq(Å$^2$) | site |
| Ba1 | 0.66667 | 0.33333 | 0.70511(2) | 1 | 0.00363(6) | 6c |
| Ba2 | 0.00000 | 0.00000 | 0.78698(2) | 1 | 0.00288(5) | 6c |
| Nb1 | 0.66667 | 0.33333 | 0.83333 | 0.75(2) | 0.00149(9) | 3b |
| Ru1 | 0.66667 | 0.33333 | 0.83333 | 0.25(2) | 0.00149(9) | 3b |
| Ru2 | 0.33333 | 0.66667 | 0.66667 | 0.88(1) | 0.00176(6) | 3a |
| Nb2 | 0.33333 | 0.66667 | 0.66667 | 0.12(2) | 0.00176(6) | 3a |
| Ru3 | 0.33333 | 0.66667 | 0.75499(2) | 0.96(1) | 0.00202(5) | 6c |
| Nb3 | 0.33333 | 0.66667 | 0.75499(2) | 0.04(1) | 0.00202(5) | 6c |
| O6 | 0.1763(2) | 0.3530(2) | 0.71080(4) | 1 | 0.0044(1) | 18f |
| O7 | 0.5028(2) | 0.4974(2) | 0.79386(4) | 1 | 0.0045(2) | 18f |

**Supplementary Table 3.** Structural parameters for single crystal $Ba_4Nb_{1.16}Ru_{2.84}O_{12}$(**$Nb_{1.16}$**) under space group *R-3* (No. 148) at 100 K. Data obtained from single crystal x-ray diffraction.

| T = 100 K | $a = b = 5.7389(2)$ Å, $c = 28.562(1)$ Å, $Z = 3$, V = 814.67(5) Å$^3$. The agreement factor $R_1 = 1.39\%$ was achieved by using 1156 unique reflections with $I > 4\sigma$ and a resolution of $d_{min} = 0.55$ Å. Anisotropic atomic displacement parameters were used for all elements. | | | | | |
|---|---|---|---|---|---|---|
| | x | y | z | Occupancy | Ueq(Å$^2$) | site |
| Ba1 | 0.66667 | 0.33333 | 0.70476(2) | 1 | 0.00363(4) | 6c |
| Ba2 | 0.00000 | 0.00000 | 0.78668(2) | 1 | 0.00304(4) | 6c |
| Nb1 | 0.66667 | 0.33333 | 0.83333 | 0.93(2) | 0.00148(8) | 3b |
| Ru1 | 0.66667 | 0.33333 | 0.83333 | 0.07(2) | 0.00148(8) | 3b |
| Ru2 | 0.33333 | 0.66667 | 0.66667 | 0.91(2) | 0.00193(7) | 3a |
| Nb2 | 0.33333 | 0.66667 | 0.66667 | 0.09(2) | 0.00193(7) | 3a |
| Ru3 | 0.33333 | 0.66667 | 0.75486(2) | 0.93(1) | 0.00206(6) | 6c |
| Nb3 | 0.33333 | 0.66667 | 0.75486(2) | 0.07(1) | 0.00206(6) | 6c |
| O6 | 0.1764(2) | 0.3529(2) | 0.71070(4) | 1 | 0.0046(2) | 18f |
| O7 | 0.5027(2) | 0.4974(2) | 0.79384(4) | 1 | 0.0042(2) | 18f |

## III. Remarks on the Trimer Lattice

The $Ba_4Nb_{1-x}Ru_{3+x}O_{12}$ series forms in a trimer lattice that consists of three face-sharing metal-oxygen octahedra (**Fig.1a**). A trimer lattice often behaves unconventionally because the internal degrees of freedom between the three metal ions of the trimer octahedra provide an extra, decisive interaction among other fundamental interactions (e.g., Coulomb and spin-orbit-interactions) that dictates physical properties. It has become increasingly clear that heavy (high atomic number) trimer lattices promise a unique pathway for discoveries of new quantum states absent in materials with other types of lattices, such as triangular and perovskite lattices. The 4*d*-electron trimer lattice is a perfect example.

The trimer lattice adopts a rhombohedral structure with the R-3 space group (No. 148), in which a Nb-O monomer separates trimer layers along the *c* axis (**Fig.1a**). Depending on Nb concentration, the system exhibits both HFSM and adjacent QSL phases. The former corresponds to $Ba_4Nb_{0.81}Ru_{3.19}O_{12}$ or $Nb_{0.81}$ and the latter $Ba_4Nb_{1.16}Ru_{2.84}O_{12}$ or $Nb_{1.16}$ (**Fig.1a**).

**IV. Supplemental Remarks on AC Susceptibility and Resistivity at the Milli-Kelvin Range**

For completeness, we show the temperature dependence of the AC magnetic susceptibility of the HFSM in the mK range as **Supplementary Fig.1.** As expected, it shows featureless responses under field, consistent with our assertion that the low-temperature heat capacity anomaly is not reflected in charge or magnetic transport channels.

The QSL becomes very insulating at low temperatures, and the resistivity reaches up to $10^5$ Ω-cm near 5 K (see Fig.1 in Ref.1). As such, resistivity measurements of the QSL in the mK range becomes impossible because the insulating sample will act as a large heater, preventing the system from approaching temperatures below 2 K.

**V. Supplemental Remarks on Entropy Change**

Temperature-dependent entropy difference, ΔS(T), between 14 T and 0 T for the QSL is calculated as ΔS(T) = $\int_{T}^{0.46K}[C(14Tesla) - C(0)]/T'dT'$, where T is the argument of ΔS, and $T'$ is the integration variable, shown in Fig. 1d in the main text. The upper bound is fixed at 0.46 K because for T < 0.46 K, C(14 T)/T > C(0 T)/T, indicating that the magnetic field enhances low-energy excitations in this range.

For further clarification, ΔS(T) is equivalent to the shaded region in the $C/T - T$ plot in Supplemental Fig.1.

**VI. Supplemental Remarks on Magnetoresistance**

Both $\rho_a$ and $\rho_c$ at higher temperatures (e.g, 5 K) for the QSL are a strong function of field-orientation, and exhibit a strong oscillatory behavior with as a function of the angle between H and the applied current I (**Supplementary Fig.2**). The magneto-resistivity ratio, defined by $\Delta\rho/\rho(0) = [\rho(H)-\rho(0)]/\rho(0)$, can be as high as 60% at 9 T. It is remarkable that $\Delta\rho/\rho(0)$ is predominantly positive for the *a*-axis $\rho_a$ and negative for the *c*-axis $\rho_c$ (**Supplementary Fig.3**). Such a giant, anisotropic oscillatory magnetoresistance suggests an orbital quantum interference in the variable range hopping regime, which is interesting in its own right (note no long-range magnetic order down to 50 mK). Nevertheless, the contrasting transport behavior further highlights the spin-charge separation in the system at low temperatures.

**VII. Supplemental Remarks on Thermal Conductivity and Heat Capacity**

$\Delta\kappa_a/\kappa_a(0T) = [\kappa_a(14T) - \kappa_a(0T)]/\kappa_a(0T)$ is considerably greater in the QSL than in the HFSM. This is consistent with there being more spinons as heat carriers in the QSL than in the HFSM (consistent also with the larger C and κ in the QSL), which naturally makes the QSL a much better thermal conductor than the HFSM. Indeed, ΔC/C(0) is larger in the QSL than in the HFSM. These observations provide an additional, key testament to the crucial role of spinons in the behavior of the trimer systems.

Note that the spinons also dominate C(T) and κ(T) in the HFSM where the Wiedemann-Franz law is strongly violated, consistent with the spin-charge separation.

**Supplementary Figure Captions**

**Fig.1. Field induced entropy increase:** Upper Panel: Temperature dependence of C(0T)/T and C(14T)/T. Lower Panel: Zoomed-in temperature dependence of C(0T)/T and C(14T)/T. Note that at $T < 0.46$ K, C(14T) > C(0T).

**Fig. 2. Low temperature magnetic behavior:** The temperature dependence of the *a*-axis AC susceptibility $\chi'_a$ for the HFSM at representative H for 50 mK $\leq$ T $\leq$ 1 K.

**Fig. 3. The angular dependence of the magneto-resistivity ratio**, defined by $\Delta\rho/\rho(0) = [\rho(H)-\rho(0)]/\rho(0)$, of the QSL at T =5 K: **a** and **b**, for the current I || *a*-axis and **c** and **d**, for I || *c*-axis. Insets: Definition of angle q and configurations of measurements.

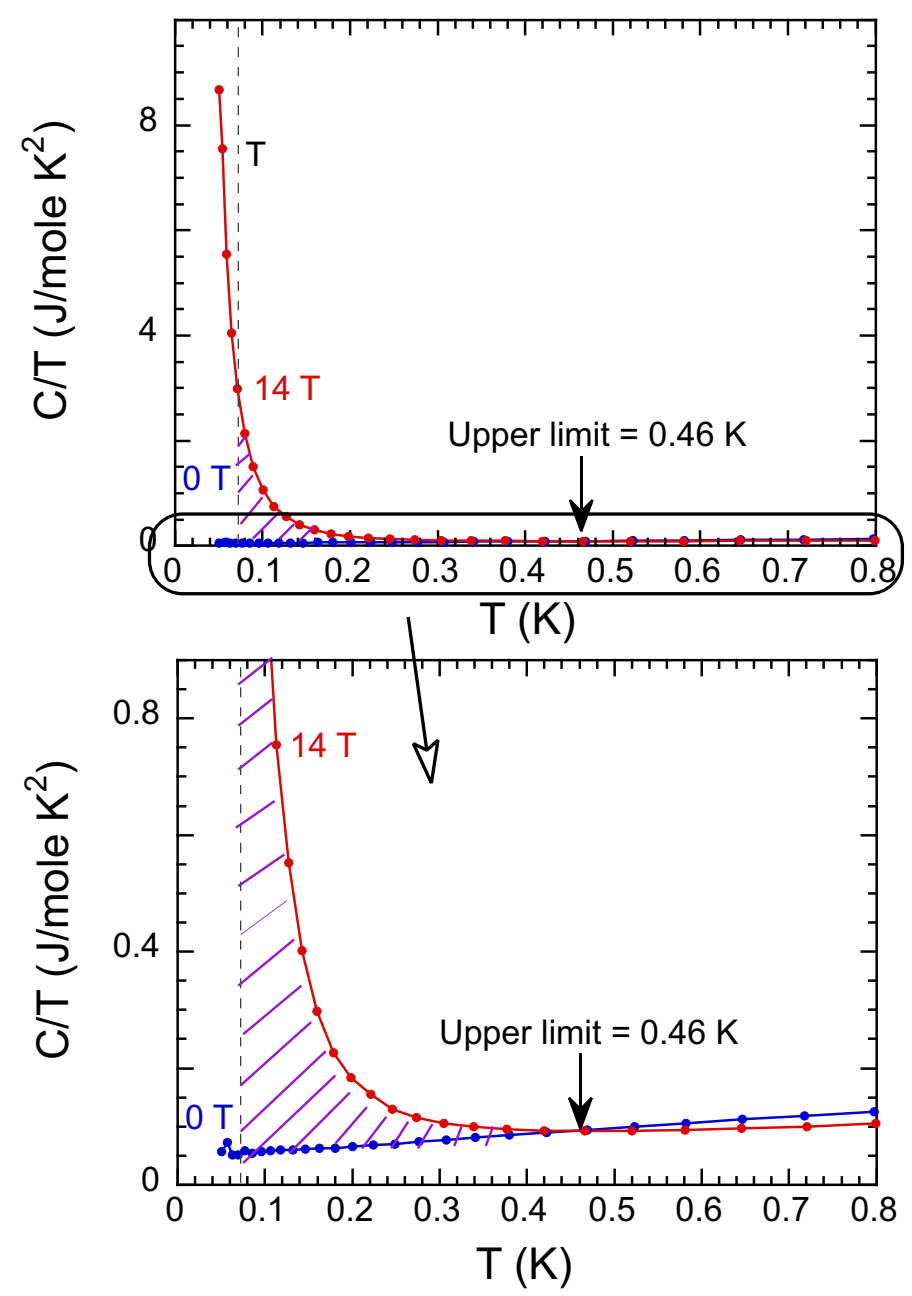

Supplementary Figure 1

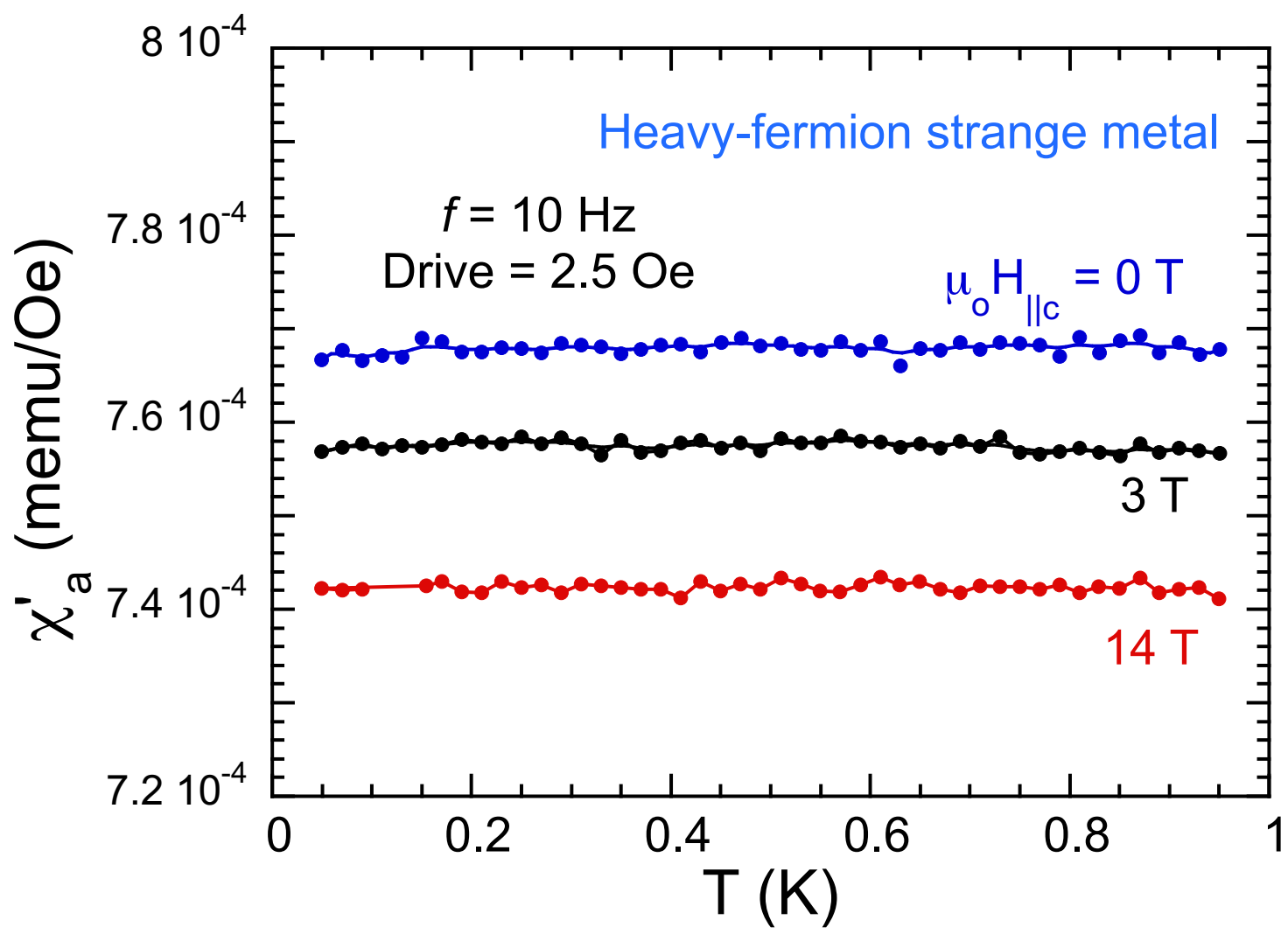

Supplementary Figure 2

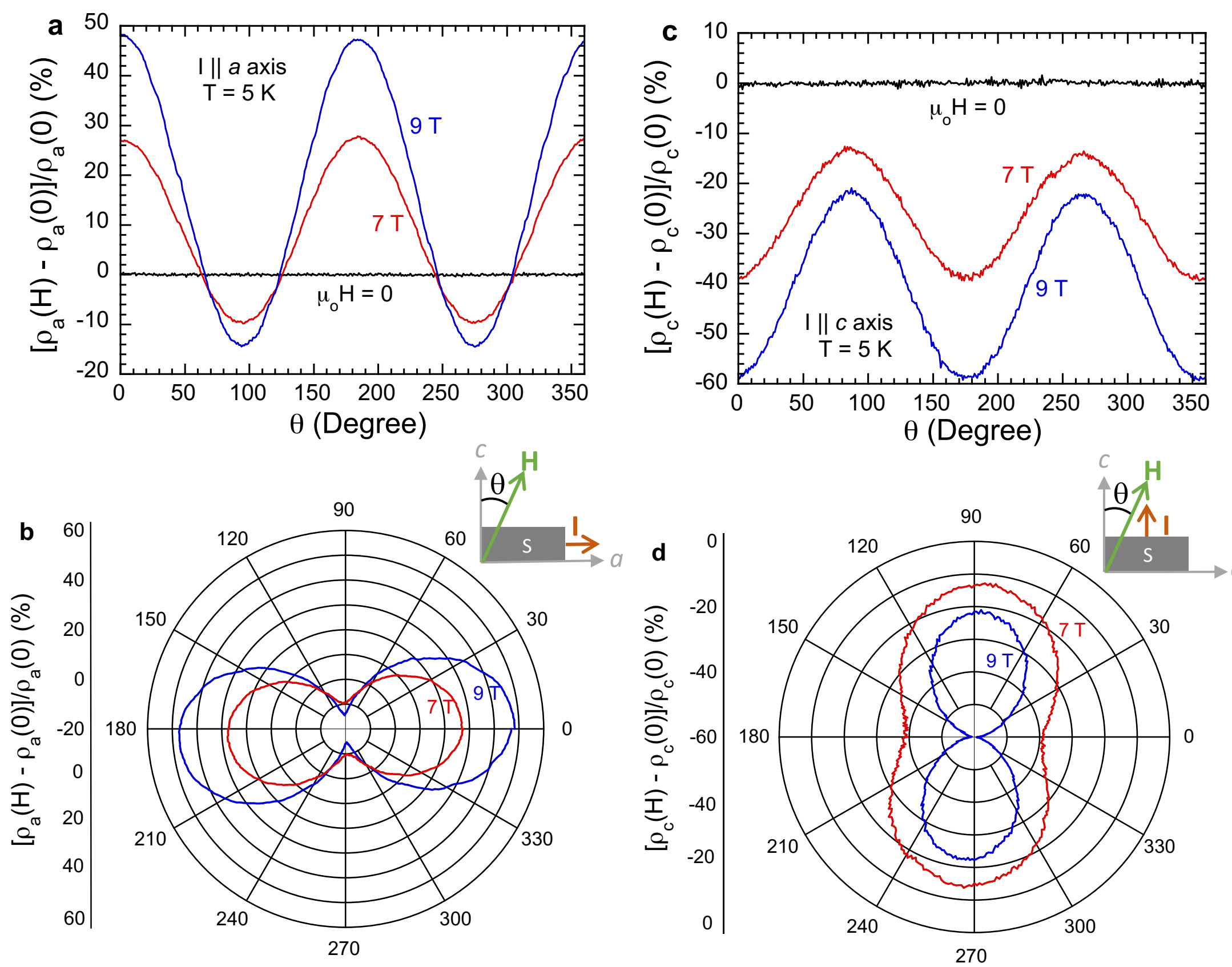

Supplementary Figure 3